# Centrality dependence of K*(892)$^0$ and φ(1020) production at LHC

Inam-ul Bashir* and Saeed Uddin

*Department of Physics, Jamia Millia Islamia (Central University) New-Delhi, India*

**Abstract**

We study the centrality dependence of the mid-rapidity (|y|<0.5) yield (dN/dy) and transverse momentum distributions of K*(892)$^0$ and φ(1020) resonances produced in Pb+Pb collisions at $\sqrt{s_{NN}}$ = 2.76 TeV. The mid-rapidity density (dN/dy) and the shape of the transverse momentum spectra are well reproduced by our earlier proposed Unified Statistical Thermal Freeze-out Model (USTFM) which incorporates the effects of both longitudinal as well as transverse hydrodynamic flow. The freeze-out properties in terms of kinetic freeze-out temperature and transverse flow velocity parameter are extracted from the model fits to the ALICE data. The extracted kinetic freeze-out temperature is found to increase with decrease in centrality while as the transverse flow velocity parameter shows a mild decrease towards peripheral collisions. Moreover the centrality dependence of the mid-rapidity system size at freeze-out has also been studied in terms of transverse radius parameter.

*inamhep@gmail.com*

## Introduction

The particle spectra of identified hadrons obtained in high energy heavy ion collisions have attracted physicists for many decades due to their fundamental nature and simplicity. The existence of QGP signals in nucleus-nucleus collisions at different nucleon-nucleon center-of-mass ($\sqrt{s_{NN}}$) energies have been studied from few GeV at the Alternating Gradient Synchrotron (AGS) to few thousand GeV recently attained in Pb-Pb and p-p collisions at CERN-LHC.

Statistical thermal models have successfully reproduced the essential features of particle production in heavy-ion collisions [1] as well as in many types of elementary collisions [2]. Furthermore, the statistical hadronization model (SHM) supplemented with the hydrodynamical expansion of the matter, to a large extent, also reproduces transverse momentum spectra of different particle species [3,4]. It has also provided a very useful framework for the centrality and system-size dependence of particle production [5]. The transverse momentum distributions of the different particles contain two components, one random and one collective. The random component can be identified as the one that depends on the temperature T of the system at kinetic freeze-out. The collective component, which arises from the matter density gradient from the center to the boundary of the fireball created in high-energy nuclear collisions, is generated by collective flow in the transverse direction and is characterized by its velocity $\beta_T$.

Assuming that the system attains thermo-chemical equilibrium at freeze-out, the momentum distributions of hadrons, emitted from within an expanding fireball, are characterized by the Lorentz-invariant Cooper-Frye formula [6]

$$E\frac{d^3n}{d^3P} = \frac{g}{(2\pi)^3} \int f\left(\frac{p^\mu u^\mu}{T}, \lambda\right) p^\mu d\Sigma_\mu, \qquad (1)$$

Where $\Sigma_f$ represents a 3-dimensional freeze-out hyper-surface and 'g = 2J+1' is the degree of degeneracy of the expanding relativistic hadronic gas. The transverse flow velocity of a particle at a distance *r* from the center of the emission source, as a function of the surface velocity $\beta_T^s$ of the expanding cylinder, is parametrized as [7] $\beta_T(r) = \beta_T^s (r/R)^n$, where *n* is a velocity profile index. The transverse radius of fireball *R* is parameterized as [7-10] $R = r_0 \exp(-\frac{z^2}{\sigma^2})$ ... (2) where $r_0$ is a free parameter which determines the transverse size of the fireball at mid-rapidity. The surface transverse expansion velocity $\beta_T^s$ is fixed in the model by using the following parameterization [7-10]

$$\beta_T^s = \beta_T^0 \sqrt{1 - \beta_z^2} \qquad (3)$$

where $\beta_T^0$ is a model parameter which gives us the surface transverse expansion velocity at mid-rapidity. Also we have chosen the chemical potential in a way so as to make it rapidity dependant [7-10], $\mu_B = a + by_0^2$ ... (4) where $y_0 = cz$ is the rapidity of the expanding hadronic fluid element along the beam axis (z-axis) and c is the constant of proportionality. This simple linear type dependence of $y_0$ on z ensures that under the transformation $z \rightarrow -z$, we will have $y_0 \rightarrow -y_0$, thereby preserving the symmetry of the hadronic fluid flow about z=0 along the rapidity axis in the centre of mass frame of the colliding nuclei. In our analysis, we have also taken into account the contributions from different heavier decay resonances [8-10]. The rest of the details of our model can be found from [7-10].

Hydrodynamic models [11, 12] that include radial flow successfully describe the measured $p_T$ distributions in Au+Au collisions at $\sqrt{s_{NN}}$ = 130 GeV [13]. The $p_T$ spectra of identified charged hadrons below 2 GeV in central collisions have been well reproduced in some models by two simple parameters: transverse flow velocity $\beta_T$ and thermal freeze-out temperature T under the assumption of thermalization. Some statistical thermal models have successfully described the particle abundances at low $p_T$ [14]. It has been shown earlier [8, 15] that our model can simultaneously explain the rapidity and transverse momentum distributions of hadrons in Au-Au collisions at RHIC energies. Also we have employed this model to successfully reproduce the transverse momentum distributions of hadrons produced in the *central* Pb + Pb collisions at $\sqrt{s_{NN}}$ = 2.76 TeV at LHC [10]. In these papers, evidences of the sequential/systematic freeze-out of the different particle species were found. However none of the two resonance particle species ($K^*(892)^0$ and $\varphi(1020)$) were studied in these papers. It will therefore be interesting to study the freeze-out properties of these particles to know whether these particles also follow the scenario of sequential freeze-out or not in Pb+Pb collisions at $\sqrt{s_{NN}}$ = 2.76 TeV. Moreover, the $\varphi(1020)$ meson serves as a very good "thermometer" of the system. This is because its interaction with the hadronic environment is believed to be negligible [16]. On the other hand, Alvarez – Ruso and Koch [17] found that the mean free path for the φ(1020) in nuclear media is smaller than usually estimated. Ishikawa et al. [18] have obtained a value of around 35 mb for the cross section between a φ(1020) meson and a nucleon, which is larger than the previous expectations. Thus on the whole the exact value for the interaction cross section of φ(1020) meson is still not clear. If the interaction cross section of φ(1020) meson is assumed to be negligible, then it may result in the early decoupling of the φ(1020) meson from the rest of the system and hence its contribution to the collective flow should be smaller than the other hadrons of similar mass. In other words its freeze-out temperature should be higher than the other similar mass hadrons. Thus it will be interesting to study the freeze-out properties of these φ(1020) mesons in order to see whether these mesons develop the significant collective effects or not. This in turn will help us to get an idea of the hadronic rescattering cross-section of these φ(1020) mesons. Also, it receive almost no contribution from resonance decays, hence its spectrum directly reflects the thermal and hydrodynamical conditions at freeze-out.

We therefore, in this paper, have studied the freeze-out properties of $K^*(892)^0$ and φ(1020) particle species by reproducing their transverse momentum distributions for various collision centralities at $\sqrt{s_{NN}}$ = 2.76 TeV in Pb+Pb collisions by using our earlier proposed Unified Statistical Thermal Freeze-out model. We have also studied the system size dependence of these particle production by reproducing their mid-rapidty density dN/dy as a function of collision centrality.

**Results and Discussion**.

In our analysis of the Transverse momentum spectra of $K^*(892)^0$ and φ(1020) shown in Figure 1, the best fit is obtained by minimizing the distribution of $\chi^2$ given by [19],

$$\chi^2 = \sum_i \frac{(R_i^{exp} - R_i^{theor})^2}{\epsilon_i^2} \tag{4}$$

where $R_i^{exp}$ is the measured value of the yield with its uncertainty $\epsilon_i$ and $R_i^{theor}$ is the value from the model calculations. The $\chi^2/dof$ is minimized with respect to the variables

T and $\beta_T^0$ whereas the values of $r_0$ are obtained by fitting the available mid-rapidity dN/dy yield (Figure 2) by using the standard relation as:

$$\frac{dn}{dy} = \int (E\frac{d^3n}{d^3p})dp_T \tag{5}$$

We have taken the values of *a* and *b* both to be zero for both the particles under the assumption of a baryon symmetric matter expected to be formed under the condition of a high degree of nuclear transparency in the nucleus-nucleus collisions at LHC energy that is, an ideal Bjorken picture [10]. Also the value of $\sigma$ is taken to be unity, because it was found to be insensitive on the mid-rapidty density (dN/dy). The ALICE experimental data points taken from [20] are shown by colored shapes whereas our model results are shown by black curves and shapes. The error bars in all the cases represent the statistical errors only. A fairly good agreement is seen between the experimental data points and the model results. The freeze-out conditions in terms of kinetic freeze-out temperature T and the transverse flow velocity parameter $\beta_T^0$ are extracted from the fits of the transverse momentum spectra of the particles, as a function of collision centrality and are tabulated in Table 1 along with the values of the free exponent *n* and the $\chi^2/dof$.

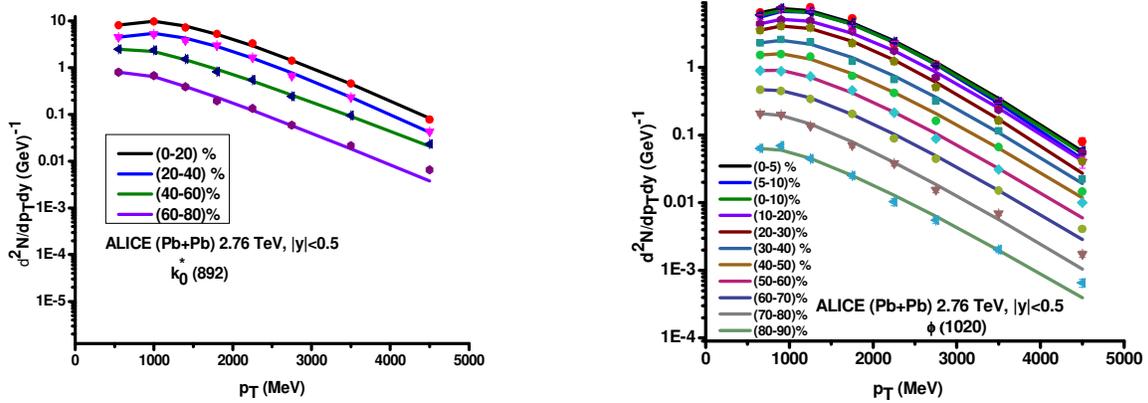

**Figure 1.** Transverse momentum spectra of $K^*(892)^0$ (left) and $\varphi(1020)$ (right) as function of collision centrality.

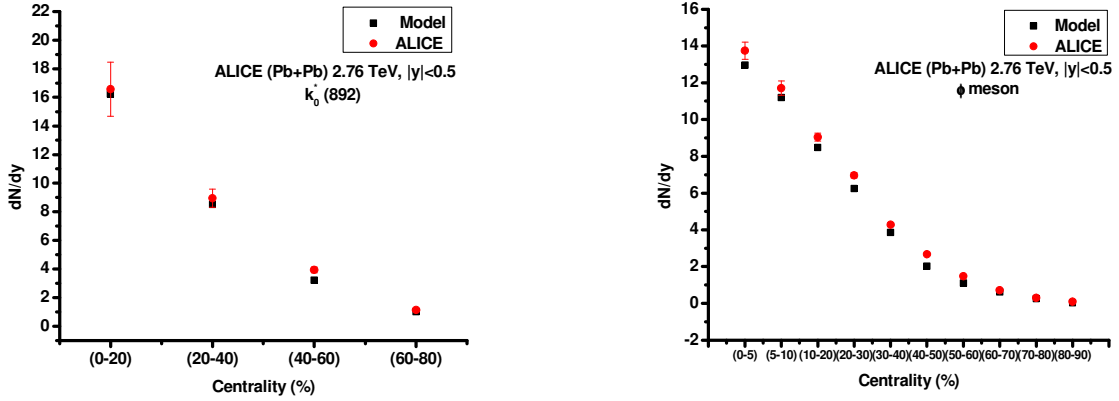

**Figure 2.** Mid-rapidity dN/dy yield of K$^*$(892)$^0$ (left) and φ(1020) (right) as function of collision centrality.

Table 1. Freeze-out parameters (T, $\beta_T^0$ and $n$) obtained from the Transverse momentum distributions of K$^*$(892)$^0$ and φ(1020) and transverse size parameter ($r_0$) obtained from the mid-rapidity yield of K$^*$(892)$^0$ and φ(1020) for different centrality classes.

| Particle | Centrality(%) | $\beta_T^0$ | T(MeV) | $n$ | $r_0$ (fm) | $\chi^2/dof$ |
|---|---|---|---|---|---|---|
| φ(1020) | (0-5) | 0.85 | 127 | 1.20 | 17.82 | 2.0 |
| | (5-10) | 0.85 | 126 | 1.21 | 17.40 | 1.0 |
| | (0-10) | 0.85 | 127 | 1.30 | 17.80 | 0.76 |
| | (10-20) | 0.85 | 129 | 1.32 | 16.31 | 2.25 |
| | (20-30) | 0.84 | 131 | 1.37 | 15.50 | 2.35 |
| | (30-40) | 0.84 | 135 | 1.44 | 14.21 | 2.12 |
| | (40-50) | 0.83 | 141 | 1.76 | 13.90 | 2.21 |
| | (50-60) | 0.82 | 149 | 1.84 | 13.0 | 2.37 |
| | (60-70) | 0.81 | 156 | 1.97 | 12.10 | 1.96 |
| | (70-80) | 0.80 | 162 | 2.13 | 10.23 | 2.0 |
| | (80-90) | 0.81 | 163 | 2.25 | 7.53 | 2.10 |
| K$^*$(892)$^0$ | (0-20) | 0.86 | 129 | 1.27 | 16.55 | 0.51 |
| | (20-40) | 0.85 | 130 | 1.29 | 14.95 | 1.05 |
| | (40-60) | 0.83 | 143 | 1.55 | 10.52 | 1.23 |
| | (60-80) | 0.80 | 160 | 2.15 | 6.21 | 2.36 |

It is observed from the Table 1 that the freeze-out temperature T and the velocity profile exponent $n$ increases towards peripheral collisions whereas the transverse flow velocity shows a mild decrease towards peripheral collisions. This is understood because in peripheral collisions the reaction volume formed is smaller and hence has a shorter life time as compared to that in central collisions. Due to this, the particles don't get enough time to develop more collective effects and freeze-out earlier as compared to those in central collisions. This kind of behavior for freeze-out parameters as a function of collision centrality has also been observed at RHIC [21, 22] as well as at LHC [23]. When compared with the results obtained in Pb+Pb collisions at LHC [10] for most central collisions, we found that the freeze-out conditions obtained for $K^*(892)^0$ and $\varphi(1020)$ do follow the trend of the sequential freeze-out observed in [10]. The collective flow is found to be stronger at LHC than at RHIC [8]. This is because of the large amount of energy available for the particle production at LHC than at RHIC. The significant value of collective flow for $\varphi(1020)$ is an indication of the larger interaction cross section of this particle in nuclear medium. Also the transverse size of the system at mid-rapidity at freeze-out is found to decrease towards peripheral collisions as expected in view of the geometry of the collision system.

## Conclusion

The centrality dependence of mid-rapidity density (dN/dy) and the transverse momentum distributions of $\varphi(1020)$ and $K^*(892)^0$ produced in Pb+Pb collisions at $\sqrt{s_{NN}}$ = 2.76 TeV is successfully reproduced by using a Unified Statistical thermal freeze-out Model. A good agreement is seen between the model results and the experimental data points. The freeze-out conditions in terms of kinetic freeze-out temperature T and transverse surface expansion velocity coefficient $\beta_T^0$ are extracted from the fits of transverse momentum spectra at different centrality classes while as the values of transverse radius parameter $r_0$ are extracted from the mid-rapidity dN/dy yields of these mesons for different collision centralities. The transverse flow velocity is found to decrease slowly and the thermal freeze-out temperature is found to increase significantly towards more peripheral collisions in both the cases. The freeze-out conditions of $\varphi(1020)$ and $K^*(892)^0$ are found to be consistent with the scenario of sequential freeze-out found in Pb+Pb collisions at LHC. A large value of collective flow for φ(1020) is an indication of the larger interaction cross section of this particle in nuclear medium. A smooth decrease of transverse radius parameter $r_0$ is observed towards peripheral collisions which is expected in view of the collision geometry.

## Acknowledgements

We acknowledge the financial support from University Grants Commission (UGC), New Delhi for this work.